\documentclass[aps,twocolumn,pra,superscriptaddress,amsmath,amssymb]{revtex4-1}
\usepackage{dcolumn}
\usepackage{bm}
\usepackage{amsmath}
\usepackage{txfonts}
\usepackage[T1]{fontenc}
\usepackage{xspace}
\usepackage{ulem}
\ifx\pdfoutput\undefined
\usepackage[dvipdfmx]{graphicx}
\usepackage[dvipdfmx]{hyperref}
\else
\usepackage{graphicx}
\usepackage{hyperref}
\fi
\begin{document}
\let\emph\textit

\title{
  Superlattice structure in the antiferromagnetically ordered state
  in the Hubbard model on the Ammann-Beenker tiling
}

\author{Akihisa Koga }
\affiliation{
  Department of Physics, Tokyo Institute of Technology,
  Meguro, Tokyo 152- 8551, Japan
}

\date{\today}
\begin{abstract}
  We study magnetic properties in the half-filled Hubbard model
  on the Ammann-Beenker tiling.
  First, we focus on the domain structure
  with locally eightfold rotational symmetry to
  examine the strictly localized confined states
  for the tightbinding model.
  We count the number of vertices and confined states
  in the larger domains generated by the deflation operations systematically. 
  Then, the fraction of the confined states,
  which plays an important role for magnetic properties
  in the weak coupling limit,
  is obtained as $p=1/2\tau^2$, where $\tau(=1+\sqrt{2})$ is the silver ratio.
  It is also found that the wave functions for confined states are
  densely distributed in the system
  and thereby the introduction of the Coulomb interactions immediately induces
  the finite staggered magnetizations.
  Increasing the Coulomb interactions,
  the spatial distribution of the magnetizations
  continuously changes to those of the Heisenberg model.
  We discuss crossover behavior in the perpendicular space representation
  and reveal the superlattice structure in the spatial distribution
  of the staggered magnetizations.
\end{abstract}
\maketitle

\section{Introduction}
Quasicrystal without translational symmetry
has attracted much interest since its first observation
in the Al-Mn arroy~\cite{Shechtman}.
Among them, the Au-Al-Yb arroy with Tsai-type clusters~\cite{Ishimasa_2011}
is one of the interesting examples with anomalous low temperature properties.
The quasicrystal $\rm Au_{51}Al_{34}Yb_{15}$ shows
quantum critical behavior with unusual exponents,
while the approximant $\rm Au_{51}Al_{35}Yb_{14}$ shows
heavy fermion behavior~\cite{Deguchi_2012}.
These experiments should clarify that
electron correlations play an important role in the quasicrystals.
Furthermore, the superconductivity has recently been observed
in the Al-Zn-Mn quasicrystal~\cite{Kamiya_2018},
stimulating further investigations on electron correlations
and induced ordered states in the quasiperiodic systems~\cite{Watanabe,Takemori_2015,Takemura_2015,Andrade,Shinzaki_2016,Otsuki,Sakai2017,Koga_Tsunetsugu_2017,Sakai2019,Zhang2020}.

\begin{figure}[htb]
 \centering
 \includegraphics[width=\linewidth]{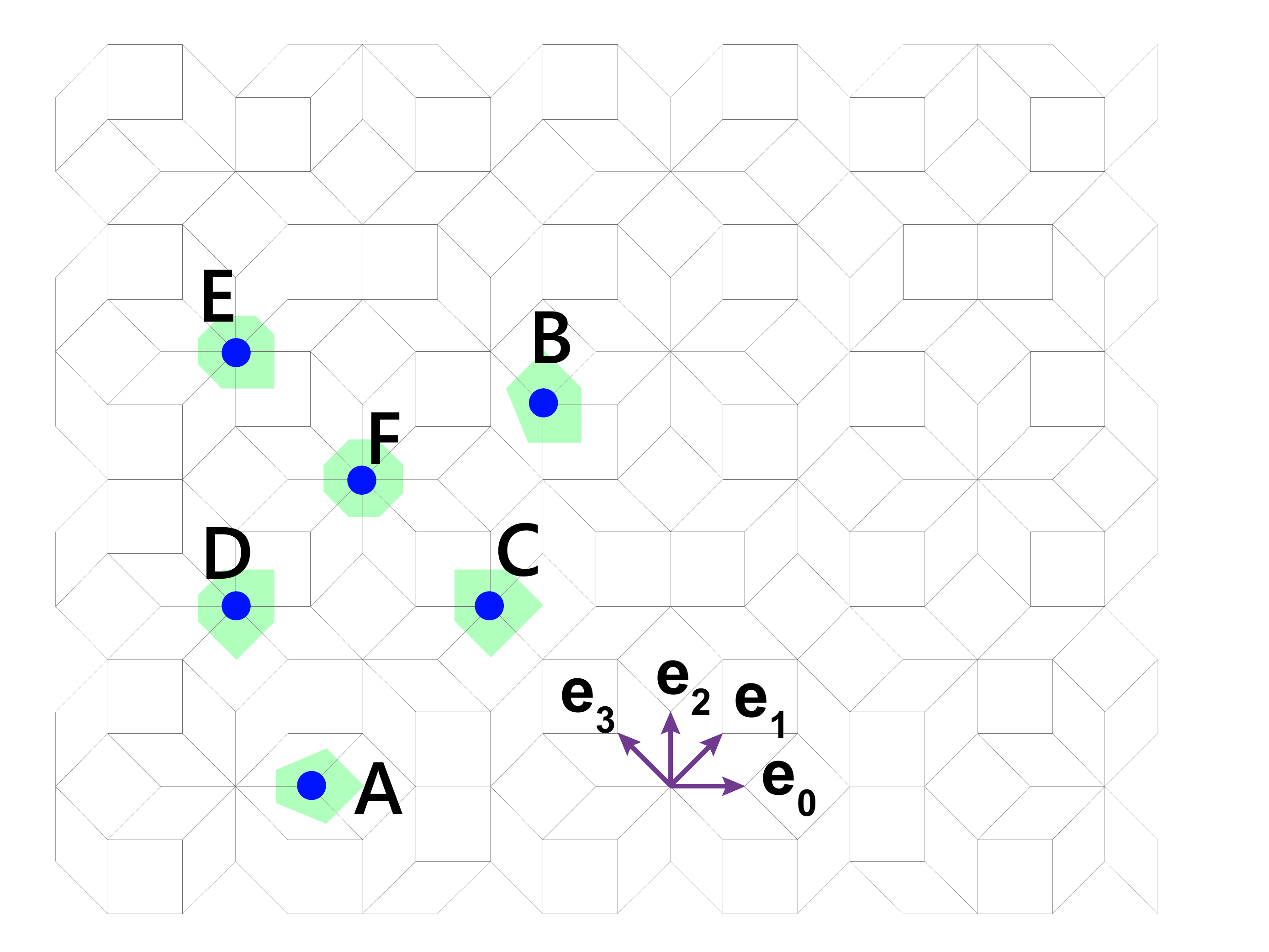}
 \caption{Ammann-Beenker tiling and six types of vertices.
   The shaded regions represent the corresponding Voronoi cells.
   ${\bf e}_0, {\bf e}_1, {\bf e}_2$, and ${\bf e}_3$ are
   projection of the fundamental
   translation vectors in four dimensions,
   ${\bf n}=(1,0,0,0), (0,1,0,0), (0,0,1,0)$, and $(0,0,0,1)$.
 }
 \label{ABlattice}
\end{figure}

Up to now, no magnetically ordered states have been found in the quasicrystals
although it has recently been observed
in the approximants $\rm Cd_6Tb$~\cite{Tamura_2010},
Au-Al-Gd~\cite{PhysRevB.93.024416} and Au-Al-Tb~\cite{PhysRevB.98.220403}.
In contrast to the experiments,
there are many theoretical works for the spontaneously symmetry breaking states
on the two-dimensional quasiperiodic lattices.
Among them, the system on the Penrose tiling~\cite{Kohmoto,Kohmoto2,Tsunetsugu_1986,Sutherland_1986,Hatakeyama_1987,Arai,Mace_2017}
has been examined, where the magnetically ordered states~\cite{Bhattacharjee_1987,Okabe_1988,Oitmaa_1990,PhysRevB.44.9271,Jagannathan_2007,Koga_Tsunetsugu_2017},
superconductivity~\cite{Sakai2017,Sakai2019,Zhang2020},
and excitonic insulator~\cite{Inayoshi_2020} have been discussed.
The Ammann-Beenker tiling~\cite{Socolar_1989,Baake_1990} (see Fig.~\ref{ABlattice})
is another example for two dimensional quasiperiodic structures,
where the superconducting~\cite{Ara_2019} and
higher order topological states~\cite{Varjas_2019}
have recently been examined.
The magnetic instability has been discussed
in the Hubbard~\cite{Jagannathan_Schulz_1997},
Heisenberg~\cite{Wessel_Jagannathan_Haas_2003,Jagannathan_2005},
and Anderson lattice~\cite{Hartman_2016} models.
However, the system size treated is not large enough
to discuss magnetic properties inherent in the quasiperiodic lattice.
In particular, the role of the strictly localized states,
which should play a crucial role in the weak coupling limit,
has not been discussed up to now.
Therefore, it is instructive to examine the confined states and
to clarify magnetic properties in the Hubbard model with larger clusters.

In the manuscript, we study the half-filled Hubbard model
on the Ammann-Beenker tiling.
First, we focus on the macroscopically degenerate states
in the noninteracting case.
By examining the domain structures
generated by the deflation operations systematically,
we obtain the fraction of the confined states in the thermodynamic limit.
To clarify the effects of the Coulomb interactions,
we apply the real-space Hartree approximation to the system
and calculate the local magnetization at each site.
We reveal that the superlattice structure appears in the weak coupling case.
Mapping the spatial distribution of the magnetization to the perpendicular space,
we also discuss the crossover in the antiferromagnetically ordered state.

The paper is organized as follows. 
In. Sec.~\ref{model},
we introduce the half-filled Hubbard model on the Ammann-Beenker tiling.
In. Sec.~\ref{conf}, we study the confined states with $E=0$,
which should play an important role for magnetic properties
in the weak coupling limit.
Counting the number of the confined states in the domains systematically,
we exactly obtain their fraction.
We discuss how the antiferromagnetically ordered state
is realized in the Hubbard model in Sec.~\ref{results}.
The crossover behavior in the ordered state is addressed,
by mapping the spatial distribution of the magnetization
to the perpendicular space.
A summary is given in the last section.

\section{Model and Hamiltonian}\label{model}
We study the Hubbard model on the Ammann-Beenker tiling,
which should be given by the following Hamiltonian,
\begin{eqnarray}
  H&=&-t\sum_{(ij)\sigma}\left(c_{i\sigma}^\dag c_{j\sigma}+h.c.\right)
  +\sum_i U\left(n_{i\uparrow}-\frac{1}{2}\right)
  \left(n_{i\downarrow}-\frac{1}{2}\right),\label{H}
\end{eqnarray}
where $c_{i\sigma} (c_{i\sigma}^\dag)$ annihilates (creates) an electron
with spin $\sigma(=\uparrow, \downarrow)$ at the $i$th site and
$n_{i\sigma}=c_{i\sigma}^\dag c_{i\sigma}$.
$t$ is the transfer integral and $U$ is the onsite Coulomb interaction.
Since the Hubbard model on the Ammann-Beenker tiling is bipartite,
the chemical potential is always $\mu=0$
when the electron density is fixed to be half filling.

The Ammann-Beenker tiling is composed of squares and rhombuses,
which is schematically shown in Fig.~\ref{ABlattice}.
There exist six types of vertices.
In the manuscript, the vertices are denoted as A, B, $\cdots$, and F
for the coordination number 3, 4, $\cdots$, and 8, respectively.
Since the vertex lattice is bipartite,
it is naively expected that the introduction of the Coulomb interactions induces
the magnetically ordered state with the staggered moments.
According to the Lieb's theorem~\cite{Lieb},
the half-filled Hubbard model on the bipartite lattice has
the total spin $S_{tot}=\frac{1}{2}|N_A-N_B|$ in the ground state,
where $N_A$ and $N_B$ are the numbers of sites in the A and B sublattices.
Therefore, the imbalance in their numbers
yields the ferrimagnetically ordered state
{\it e.g.} Lieb lattice~\cite{Noda}.
In our model, one can prove that the antiferromagnetically ordered state
is realized without uniform magnetizations, 
considering the deflation rule. 
\begin{figure}[htb]
 \centering
 \includegraphics[width=\linewidth]{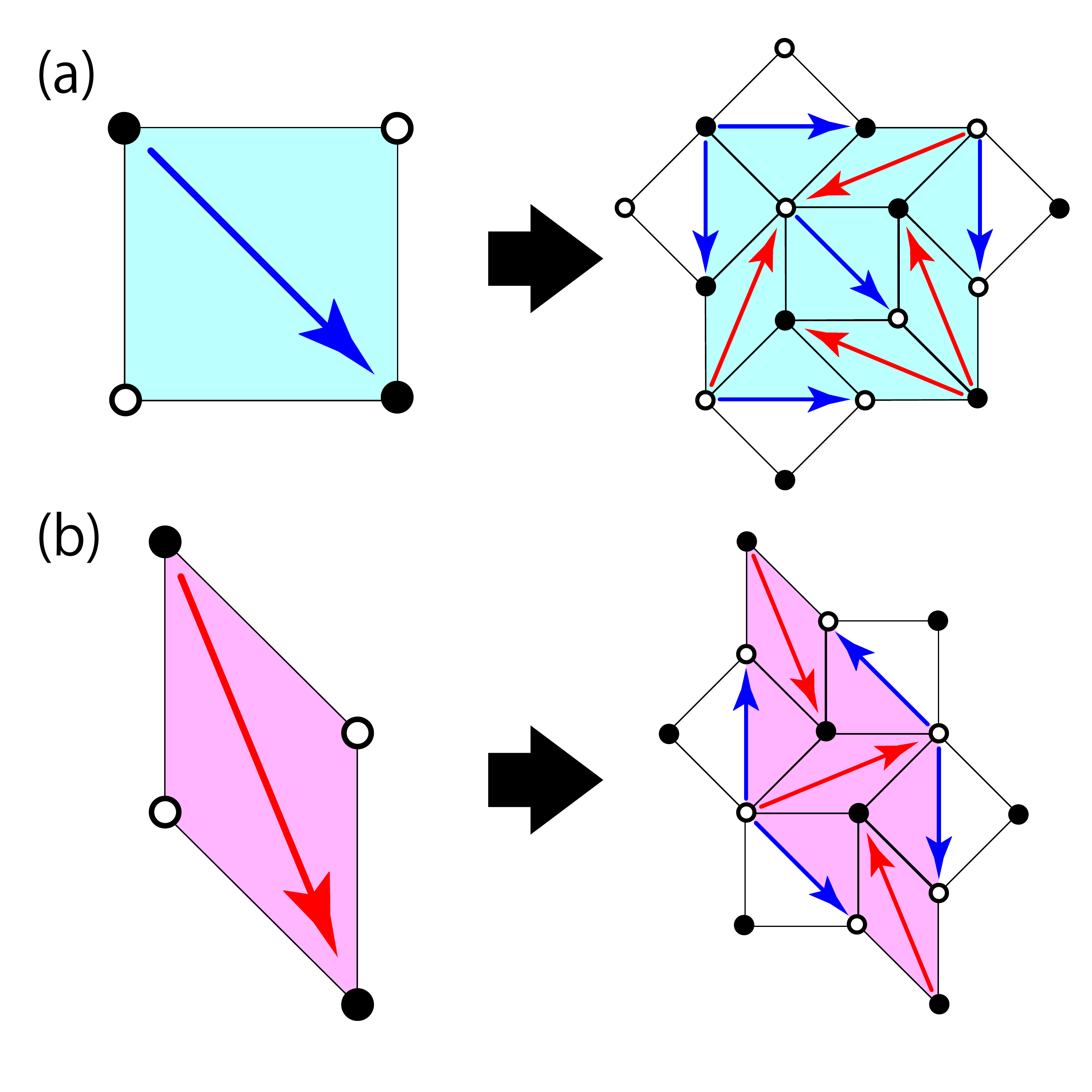}
 \caption{Deflation rule for directed square (a) and rhombus (b)
   in the Ammann-Beenker tiling.
   Open and solid circles at the corners represent
   the distinct sublattices (see text).
 }
 \label{def}
\end{figure}
Figure~\ref{def} shows the deflation rule for the directed squares and rhombuses,
where the open and solid circles at the corners
represent the distinct sublattices.
By applying the deflation operations to the squares and rhombuses,
their numbers are changed as
\begin{eqnarray}
  S_\sigma&\rightarrow& S_\sigma +2S_{\bar{\sigma}}+2R_\sigma+2R_{\bar{\sigma}},\label{S}\\
  R_\sigma&\rightarrow& 2S_{\bar{\sigma}}+2R_\sigma+R_{\bar{\sigma}},\label{R}
\end{eqnarray}
where $S_\sigma$ ($R_\sigma$) is the number of the squares (rhombuses)
with spin $\sigma$
where two spins connected by the arrows are $\sigma$
and the other spins are $\bar{\sigma}$.
It is known that in the thermodynamic limit,
the numbers of squares and rhombuses $\tau^2$ times increase
for each deflation process and $S/R=1/\sqrt{2}$,
where $\tau(=1+\sqrt{2})$ is the silver ratio~\cite{Socolar_1989,Baake_1990}.
From the above relations (\ref{S}) and (\ref{R}),
we obtain that $S_\sigma=S/2$ and $R_\sigma=R/2$. 
Since the number of squares and rhombuses are independent of spins
in the thermodynamic limit, the vertices are also independent.
Its proof is explicitly shown in Appendix.
Then, we can say that
the antiferromagnetically ordered state without uniform magnetizations
is realized in the thermodynamic limit.

\begin{figure}[htb]
 \centering
 \includegraphics[width=\linewidth]{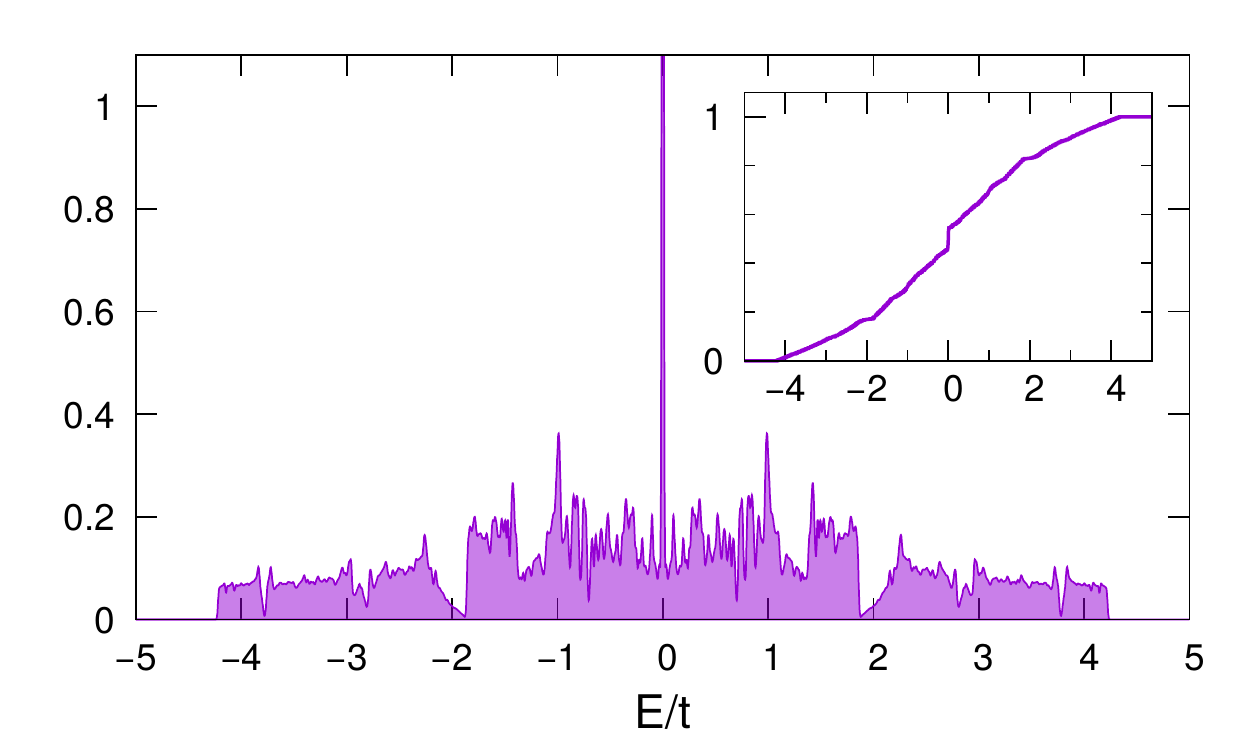}
 \caption{
   Density of states in the tightbinding model
   on the Ammann-Beenker tiling with $N=1\, 049\, 137$.
   The inset shows the integrated density of states.
 }
 \label{dos}
\end{figure}
On the other hand, the magnetization profile may not be trivial
since in the quasicrystals, each lattice site is not equivalent,
in contrast to the conventional lattice with translational symmetry.
In particular, in the weak coupling limit,
magnetic properties strongly depend on the noninteracting
density of states (DOS) at the Fermi level.
Figure~\ref{dos} shows the DOS in the tightbinding model
on the Ammann-Beenker tiling.
We find the delta-function like peak at $E=0$,
meaning the existence of the confined states.
When magnetic properties are studied at half filling,
the confined states should play an essential role
in understanding magnetic properties.
In the following section,
we focus on these macroscopically degenerate states with $E=0$.

\section{Confined states in the tightbinding model on the Ammann-Beenker tiling}
\label{conf}
In the section, we focus on the confined states 
in the tightbinding model.
As seen in Fig.~\ref{dos}, the eigenstates are
macroscopically degenerate at $E=0$, 
which means that
the corresponding states are exactly localized in certain regions.
This is similar to the model on the Penrose
tiling~\cite{Kohmoto,Arai,Koga_Tsunetsugu_2017}.
The key of the confined states is the fact that the Ammann-Beenker tiling has
the eightfold rotational symmetry.
Here, we focus on the F vertex with locally eightfold rotational symmetry,
which is closely related to the confined states, as discussed later.
Due to the matching rule of the Ammann-Beenker tiling,
there always appear eight squares and sixteen rhombuses
around each F vertex, as shown in Fig.~\ref{F}(a).
For convenience, when the local eightfold rotational symmetry is satisfied
in the domain shown in Fig.~\ref{F}(a) and is not satisfied outside,
we define this domain composed of seventeen sites (the boundary sites are excluded)
as $D_1$.
By applying the deflation operation to the domain $D_1$,
a new domain is generated, as shown in Fig.~\ref{F}(b).
If one focuses on the F vertex at the center, 
the rotational symmetry is satisfied in the domain with larger lattice sites,
which is bounded by the regular octagon shown as the dashed line in Fig.~\ref{F}(b).
This domain is denoted as $D_2$.
Repeating the deflation operations, we obtain the $D_i$ domains.
Then, we can define the F vertex at the center of the domain $D_i$
as F$_i$.
Figure~\ref{F} shows the domains $D_1$, $D_2$, and $D_3$,
where F$_1$, F$_2$, and F$_3$ vertices are located at their centers, respectively.
In the $D_3$ domain, we find sixteen $D_1$ domains with the F$_1$ vertices.
Note that there does not exist the $D_1$ domain at the center because of
its definition.
It is known that, in each deflation operation,
F$_i\; (i>1)$ vertices are generated from the F$_{i-1}$ vertices and
the F$_1$ vertices are generated from
half of the C vertices, and D and E vertices (see Fig.~\ref{F}).
Then, in the thermodynamic limit,
the fraction of the F$_i$ vertices is obtained as
\begin{eqnarray}
  p^{{\rm F}_i}=2\tau^{-(2i+3)},
\end{eqnarray}
since $p^{\rm F_1}=\left(\frac{1}{2}p^{\rm C}+p^{\rm D}+p^{\rm E}\right)/\tau^2=2\tau^{-5}$
and $p^{{\rm F}_{i+1}} = p^{{\rm F}_{i}}/\tau^2$,  
where $p^\alpha$ is the fraction of
the $\alpha(=$A, B, C, D, E, and F$)$ vertex~\cite{Baake_1990}:
$p^{\rm A}=\tau^{-1}, p^{\rm B}=2\tau^{-2}, p^{\rm C}=2\tau^{-3}, p^{\rm D}=2\tau^{-4},
p^{\rm E}=\tau^{-5}, p^{\rm F}=\tau^{-4}$.
Since the F$_i$ vertex is defined as the center vertex of the domain $D_i$,
the fraction of the domain $D_i$ is given as $p_i=p^{{\rm F}_i}$.

\begin{widetext}
  
  \begin{center} 
    \begin{figure}[htb]
      \includegraphics[width=\linewidth]{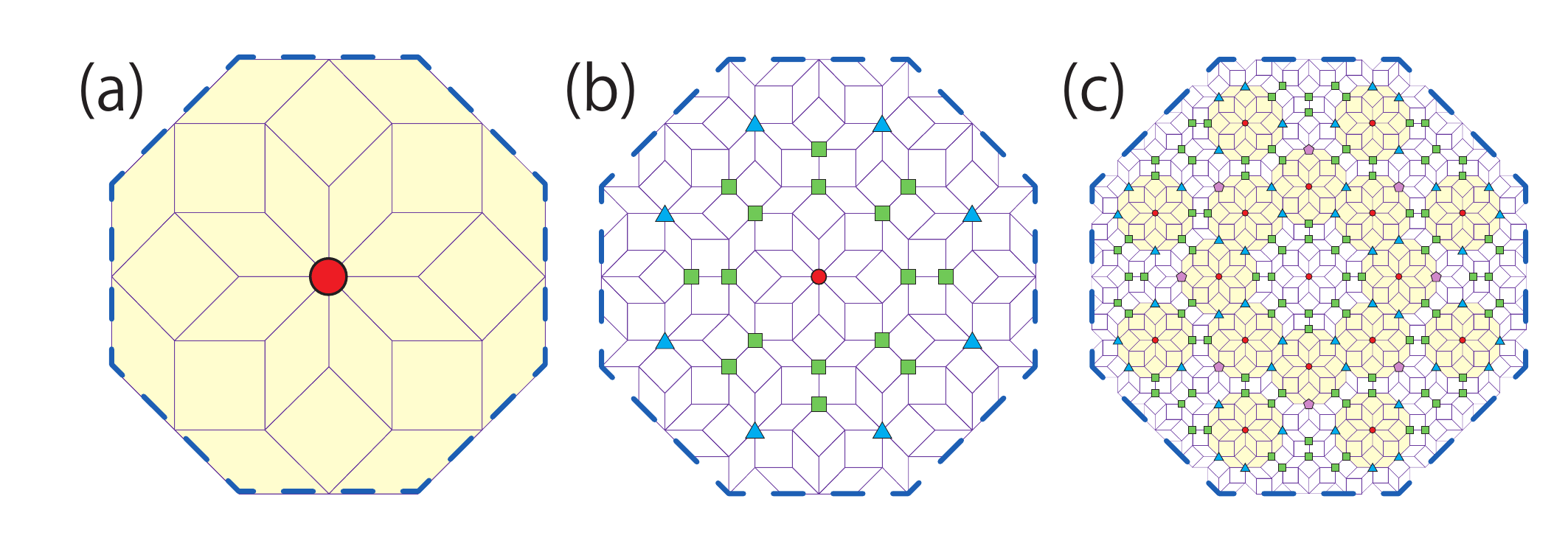}
      \caption{(a), (b), and (c) are the domains $D_1$, $D_2$, and $D_3$, respectively.
        Solid triangles, squares, pentagons, and circles represent
        C, D, E, and F vertices.
        Shaded regions represent the $D_1$ domains.
        For clarify, the lattice constant are not rescaled (see Fig.~\ref{def}).  
      }
      \label{F}
    \end{figure}
  \end{center}
%
%
\begin{center}
  \begin{table}
    \caption{Profile of each domain $D_i$.
      $p_i$ is its fraction,
      $N_i$ is the number of vertices, and
      $N^{\alpha}_i$ is the number of $\alpha$ vertices in the $i$th domain,
      where the sites on the boundary are excluded.
      $N^{tot}_i$ is the total number of the confined states,
      $N_i^{net}$ is the net number of the confined states, and
      $p_i^{conf}(=N_i^{tot}/N_i)$ is the fraction of the confined states
      in the $i$th domain (see text).
    }
    \begin{tabular}{ccr|rrrrrr|rrr}
      \hline
      \hline
$i$ & $p_i$ & $N_i$ & $N^{\rm A}_i$ & $N^{\rm B}_i$  & $N^{\rm C}_i$ & $N^{\rm D}_i$ & $N^{\rm E}_i$ & $N^{\rm F}_i$ & $N^{tot}_i$ & $N^{net}_i$ & $p_i^{conf}$ \\
      \hline
1 & $2\tau^{-5}$ &17& 8   & 8    & 0   & 0   & 0   & 1 & 2 & 2  & 0.1176\\
2 & $2\tau^{-7}$ &121& 48  & 48   & 16  & 8   & 0   & 1 & 6 & 6 & 0.0496\\
3 & $2\tau^{-9}$ &753& 312 & 272  & 96  & 48  & 8   & 17 & 44 & 12 & 0.0584\\
4 & $2\tau^{-11}$ &4\,521& 1\,872& 1\,584 & 624 & 272 & 48  & 121 & 324 & 20 & 0.0717\\
5 & $2\tau^{-13}$ &26\,673& 11\,048& 9\,232 & 3\,744 & 1\,584 & 312  & 753& 2\,110 & 30 & 0.0791\\
6 & $2\tau^{-15}$ &156\,249& 64\,720& 53\,808 & 22\,096 & 9\,232 & 1\,872  & 4\,521 & 12\,938 & 42 & 0.0828\\
7 & $2\tau^{-17}$ &912\,593& 378\,008 & 313\,616 & 129\,440 & 53\,808 & 11\,048 & 26\,673 & \,77\,112 & 56 & 0.0845\\
8 & $2\tau^{-19}$ &5\,323\,593& 2\,205\,104 & 1\,827\,888 & 756\,016 & 313\,616 & 64\,720 & 156\,249\\
9 & $2\tau^{-21}$ &31\,039\,313& 12\,856\,904 & 10\,653\,712 & 4\,410\,208 & 1\,827\,888 & 378\,008 & 912\,593 \\
10 & $2\tau^{-23}$&180\,937\,273& 74\,946\,672 & 62\,094\,384 & 25\,713\,808 & 10\,653\,712 & 2\,205\,104 & 5\,323\,593\\
11& $2\tau^{-25}$ &\,1\,054\,644\,657& \,436\,848\,120 & \,361\,912\,592 & \,149\,893\,344 & \,62\,094\,384 &\, 12\,856\,904 & \,31\,039\,313\\
      \hline
    \end{tabular}
    \label{I}
  \end{table}
\end{center}
\end{widetext}

By counting the numbers of all vertices up to the domain $D_{11}$ numerically,
we obtain the domain profile, as shown in Table~\ref{I}.
From these data, one finds relations between the number of vertices.
For examples, $N^{\rm C}_{i+1}=2N^{\rm A}_i$, $N^{\rm D}_{i+1}=N^{\rm B}_{i}$,
$N^{\rm E}_{i+2}=N^{\rm A}_{i}$, and
$N^{\rm F}_{i+1}=N^{\rm C}_{i}/2+N^{\rm D}_{i}+N^{\rm E}_{i}+N^{\rm F}_i$.
Estimating the general terms for $N^{\rm A}_i$, $N^{\rm B}_{i}$, and $N_i$ as,
\begin{eqnarray}
  N^{\rm A}_{i}&=&2\sqrt{2}\Big[(-\tau)^{1-i}-\tau^{i-1}\Big]+4\Big[\tau^{2i-1}-\tau^{1-2i}\Big],\\
  N^{\rm B}_{i}&=&8\Big[\tau^{2i-2}+\tau^{2-2i} -\delta_{i1}\Big],\\
  N_i&=&1+2\sqrt{2}\Big[(-\tau)^{-i}-\tau^i\Big]+4\Big[\tau^{2i}+\tau^{-2i}\Big],
\end{eqnarray}
we obtain the general terms for all vertices in each domain.
Namely, the domain $D_\infty$ can be regarded as the Ammann-Beenker tiling
in the thermodynamic limit and we have confirmed that
the fraction for each vertex
$p_\infty^\alpha=\displaystyle \lim_{i\rightarrow\infty} N^\alpha_i/N_i$
is reduced to the well-known value $p^\alpha$~\cite{Baake_1990}.

Now, we consider the confined states in each domain
with the eightfold rotational symmetry.
In the domain $D_1$, there are two confined states.
\begin{figure}[htb]
 \centering
 \includegraphics[width=\linewidth]{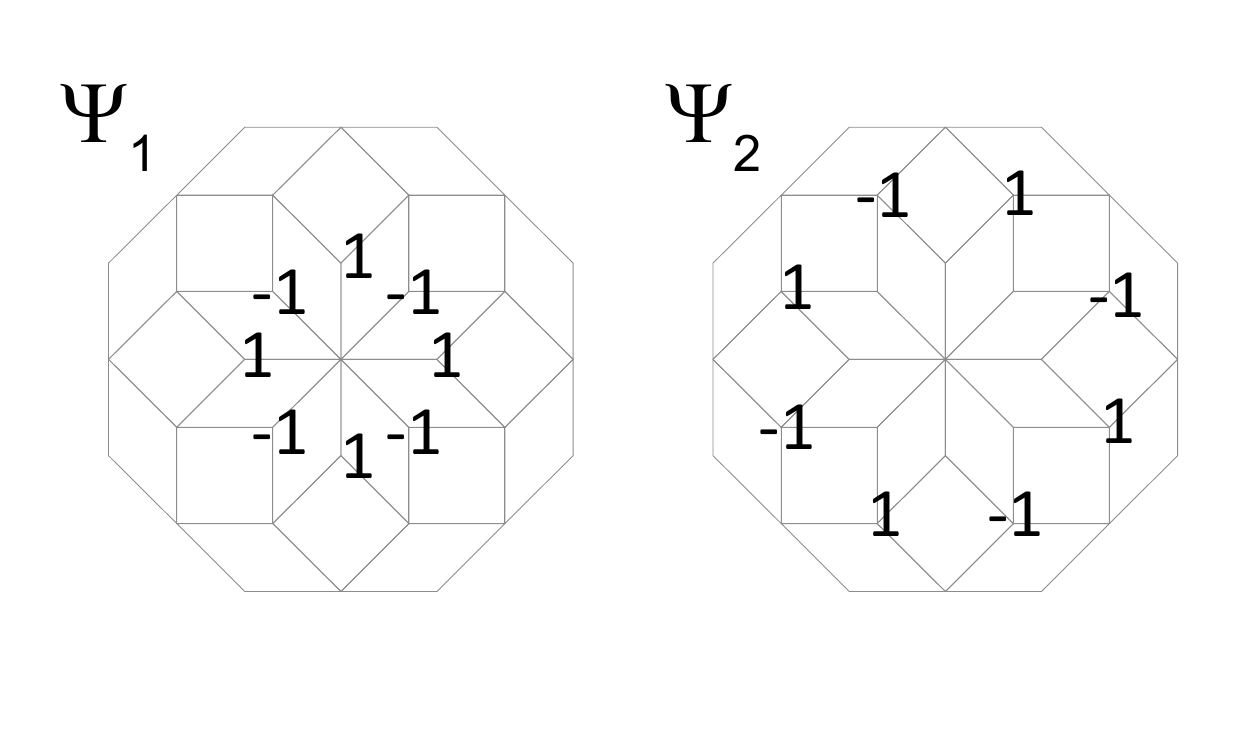}
 \caption{
   Two confined states in the domain $D_1$ for the tight-binding model
   on the Ammmann-Beenker tiling.
   The number at the vertices represent the amplitudes of confined state.
 }
 \label{conf1}
\end{figure}
Since the confined states satisfy the Sch\"odinger equation $H\Psi=0$ with $U=0$,
it is always possible to choose each eigenstate such that
it can be described by the irreducible representation of the point group $D_8$.
Table~\ref{II} shows a part of the irreducible characters
of the dihedral group $D_8$,
where there exist four one-dimensional irreducible representations.
Namely, the confined states $\Psi_1$ and $\Psi_2$,
which are schematically shown in Fig.~\ref{conf1},
are described by the irreducible representation B$_1$ and B$_2$,
and $\langle \Psi_1|\Psi_2\rangle=0$.
We wish to note that these confined states are always exact eigenstates
even when the system does not have eightfold rotational symmetry outside of
the domain $D_1$.
We also find that the amplitudes of the wave function $\Psi_1$ are finite only
in the sublattice B,
and the others are in the sublattice A
when the sublattice for the center site is regarded as the sublattice A.
This is contrast to the case in the vertex model
on the Penrose tiling~\cite{Kohmoto,Arai},
where finite amplitudes appears in one of the sublattices
in the {\it cluster} defined in Ref~\cite{Koga_Tsunetsugu_2017}.
This should induce distinct spatial distribution of the magnetization
in the weak coupling limit,
which will be discussed in the next section.
\begin{center}
  \begin{table}
    \caption{A part of the irreducible characters of the dihedral group $D_8$.
      $E$ is an identity operator, $C_8$ is a rotation operator of $\pi/4$,
      and $I_y$ is a reflection operator about the $y$ axis.}
    \begin{tabular}{crrrr}
      \hline
      \hline
            & $E$ & $C_8$  & $I_y$ & $I_yC_8$\\
      \hline
      A$_1$ &  1 &  1 &  1 & 1\\
      A$_2$ &  1 &  1 & -1 & -1\\
      B$_1$ &  1 & -1 &  1 & -1\\
      B$_2$ &  1 & -1 & -1 &  1\\
      \hline
    \end{tabular}
    \label{II}
  \end{table}
\end{center}

\begin{figure}[htb]
 \centering
 \includegraphics[width=\linewidth]{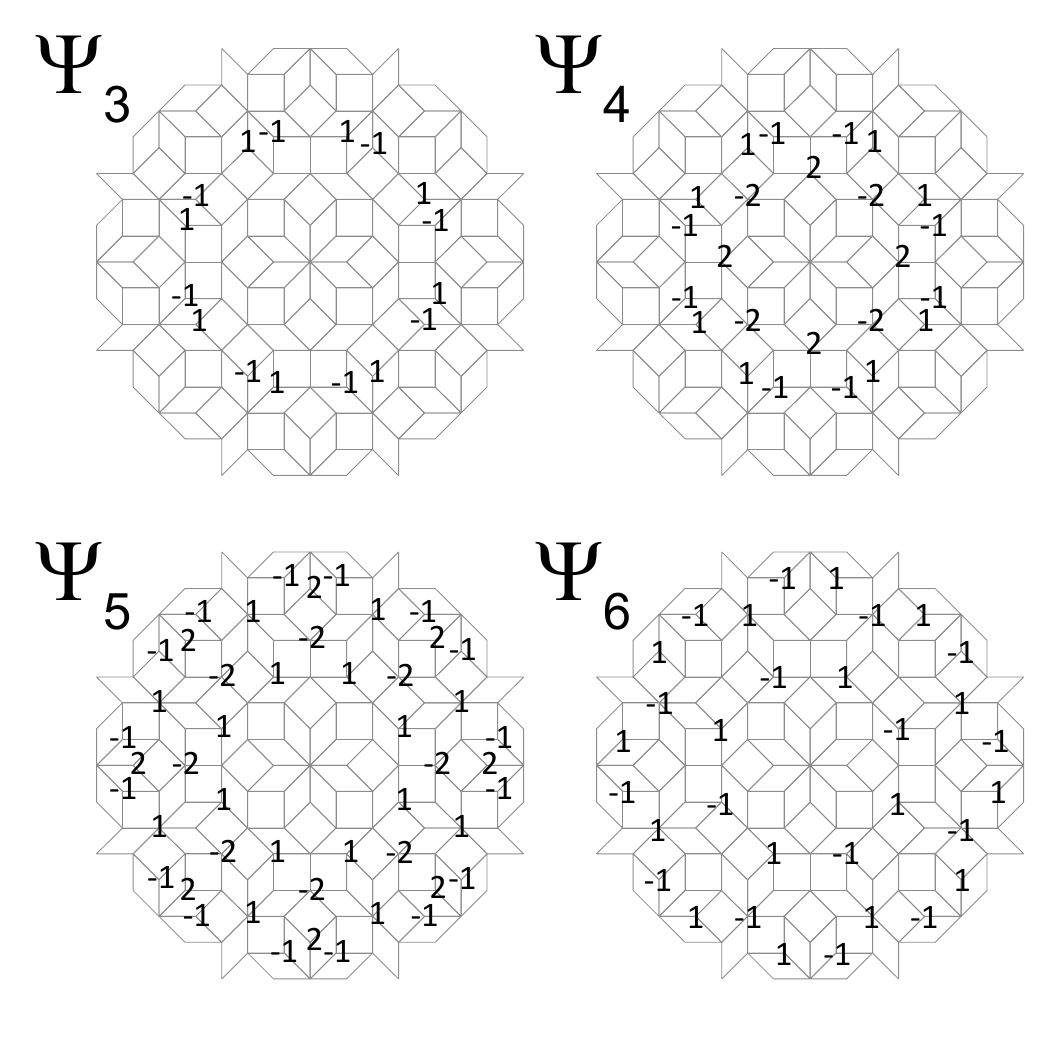}
 \caption{
   Four confined states in the domain $D_2$ for the tight-binding model
   on the Ammmann-Beenker tiling.
   The number at the vertices represent the amplitudes of confined state.
 }
 \label{conf2}
\end{figure}

\begin{figure}[htb]
 \centering
 \includegraphics[width=\linewidth]{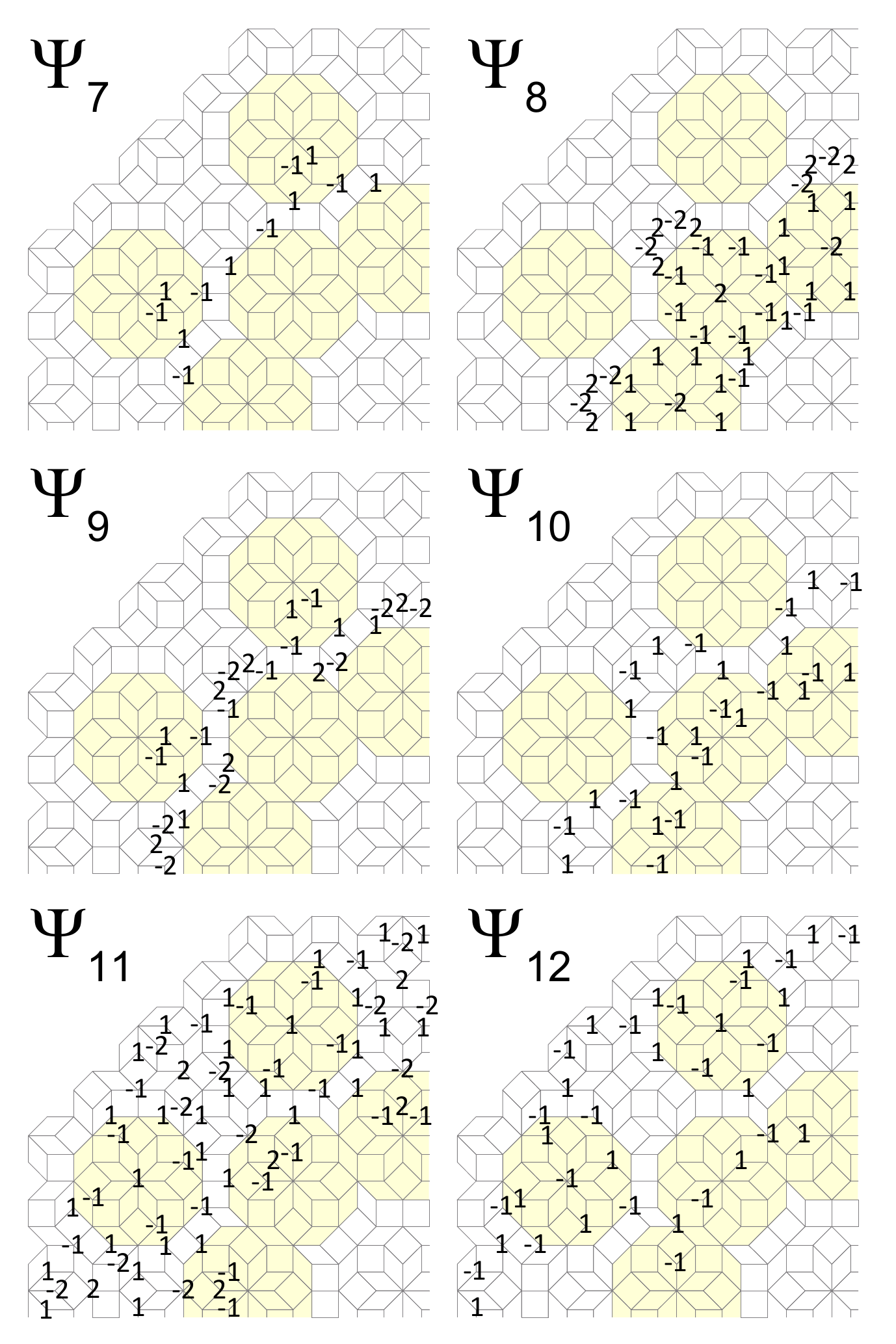}
 \caption{
   Six confined states in the domain $D_3$ for the tight-binding model
   on the Ammmann-Beenker tiling.
   For clarity, the quarter of the domain $D_3$ is shown.
   The shaded areas represent the $D_1$ domains inside of the domain $D_3$
   and the F$_1$ vertex is located in each domain $D_1$.
   The number at the vertices represent the amplitudes of confined state.
 }
 \label{conf3}
\end{figure}

In the domain $D_2$, there is the structure of the domain $D_1$ around the center.
Therefore, in the domain $D_2$,
$\Psi_1$ and $\Psi_2$ located there are the confined states.
Furthermore,
we find four confined states $\Psi_3$, $\Psi_4$, $\Psi_5$, and $\Psi_6$,
as shown in Fig.~\ref{conf2}.
It is found that these confined states are described by
the irreducible representations A$_2$, B$_1$, A$_1$, and B$_2$.
Namely, $\Psi_1$ and $\Psi_4$ ($\Psi_2$ and $\Psi_6$) are described by
the same irreducible representation B$_1$ (B$_2$), but there are no overlap
in their wave functions.
In the domain $D_3$, in addition to the six confined states shown above,
we find six confined states $\Psi_7, \Psi_8, \cdots$, and $\Psi_{12}$,
which are explicitly shown in Fig.~\ref{conf3}.
These are described by the irreducible representations
A$_2$, B$_1$, B$_1$, B$_2$, A$_1$, and B$_2$. 
We note that, in the domain $D_3$, there exist sixteen $D_1$ domains
(shown as the shaded regions in Figs.~\ref{F} and \ref{conf3}),
where two confined states $\Psi_1$ and $\Psi_2$ exist locally.
Therefore, in the domain $D_3$,
the net number of the confined states $N_3^{net}=12$,
and the total number of the confined states $N_3^{tot}=N_3^{net}+16N_1^{net}=44$,
with $N_1^{net}=2$.

To count the number of the confined states in larger domains systematically,
we perform the exact diagonalization method for the tightbinding Hamiltonian.
The results up to the domain $D_7$ are shown in Table~\ref{I}.
The net number of the confined states is evaluated
by taking into account the smaller domains, as
$N^{net}_i=N^{tot}_i-\sum_{j=1}^{i-1} N_{ij} N^{net}_j$,
where $N_{ij}$ is the number of domain $D_j$ inside of the domain $D_i$.
Namely, $N_{ij}$ satisfies the relations as
$N_{i+1,j+1}=N_{ij}$, $N_{i1}=N^{\rm C}_{i-1}/2+N^{\rm D}_{i-1}+N^{\rm E}_{i-1}$,
and $N^{\rm F}_i=\sum N_{ij}$,
where $N_i^\alpha$ is the number of the $\alpha$ vertex in the domain $D_i$.
Since the net number of confined states should be given as $N^{net}_i=i(i+1)$,
we obtain the fraction of the confined states in the tightbinding model
on the Ammann-Beenker tiling as,
\begin{eqnarray}
  p&=&\sum_i p_i N_i^{net}=\frac{1}{2\tau^2}\sim 8.579 \times 10^{-2},
\end{eqnarray}
where $p_i$ is the fraction of the $D_i$ domain.
We have also confirmed that it corresponds to the ratio
in the domain $D_\infty$, $\displaystyle p=\lim_{i\rightarrow\infty} N_i^{tot}/N_i$,
where the general term for the total number of the confined states is given as
\begin{eqnarray}
N_i^{tot}=4+2\sqrt{2}[(-\tau)^{-i}-\tau^i]+2(\tau^{2i-2}+\tau^{2-2i})+i(i+1).
\end{eqnarray}

In the following, we consider electron correlations in the Hubbard model
to discuss how the antiferromagnetically ordered state is realized
in the Ammann-Beenker tiling.
In the weak coupling limit, it is, in principle, possible
to evaluate the magnetization by means of
the Gram-Schmidt orthogonalization for the confined states at $E=0$
since their degeneracy should be lifed by
the introduction of the Coulomb interactions.
However, the confined state are densely distributed in the lattice.
Figures~\ref{conf1}, \ref{conf2}, and \ref{conf3} show that
the confined states have amplitudes in almost whole of the domain.
In addition, the amplitudes of confined states in a certain domain $D$
sometimes appear on the smaller domains inside of $D$,
where some confined states exist locally.
For example, in Fig.~\ref{conf3},
the wave function $\Psi_{11}$ has amplitudes in each domain $D_1$
(the shaded areas) with the local wave functions $\Psi_1$ and $\Psi_2$ .
Therefore, the wave functions for confined states multiply overlap in the space.
This is contrast to the Penrose-Hubbard model,
where
there exist finite number of confined states in a certain region
``{\it cluster}'' and
the seventy percents of magnetizations are exactly
obtained in the thermodynamic limit~\cite{Koga_Tsunetsugu_2017}.
\begin{figure}[htb]
 \centering
 \includegraphics[width=\linewidth]{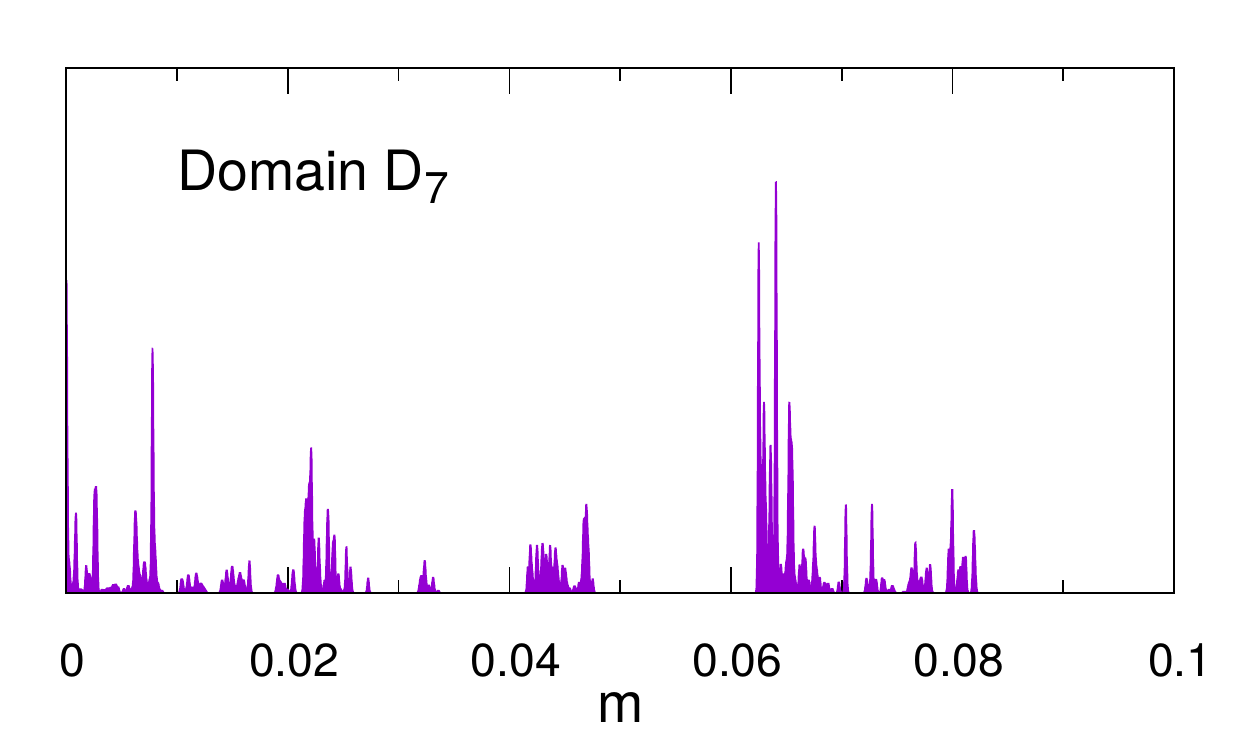}
 \caption{
   Local magnetization in the $D_7$ domain
   for the Hubbard model on the Ammann-Beenker tiling
   in the limit $U\rightarrow 0$. 
 }
 \label{dens-non}
\end{figure}
Figure~\ref{dens-non} shows the local magnetization for the $D_7$ domain,
which is obtained from 77\,112 confined states.
It is found that the distribution of the magnetization is classified into
some groups.
The group with large magnetizations is mainly contributed from
the A and B vertices around the F vertex,
which originates from the confined states $\Psi_1$ and $\Psi_2$
(see Fig.~\ref{conf1}).
In the manuscript, we apply the simple mean-field theory to the Hubbard model
to discuss magnetic properties inherent
in the Ammann-Beenker tiling~\cite{Jagannathan_Schulz_1997}.

\section{Antiferromagnetically ordered state}\label{results}
In the section, we consider the Hubbard model with finite $U$.
To study the antiferromagnetically ordered state characteristic of
the Ammann-Beenker tiling,
we make use of the real-space Hartree approximation and
the Hamiltonian (\ref{H}) is reduced to 
\begin{eqnarray}
  H_{MF}&=&-t\sum_{\langle ij\rangle \sigma}\left(c_{i\sigma}^\dag c_{j\sigma}+h.c.\right)
  +U\sum_{i\sigma}\langle n_{i\bar{\sigma}}\rangle n_{i\sigma},
\end{eqnarray}
where $\langle n_{i\sigma} \rangle$ is the expectation value of
the number of electron with spin $\sigma$ at the $i$th site.
In our calculations, we use the open boundary condition and
examine finite lattices with $N=180\,329$ and $1\,049\,137$,
where the largest domains are $D_6$ and $D_7$, respectively.
The lattices are generated by the deflation operations
to the $D_1$ domain [shown in Fig.~\ref{F}(a)], and
therefore have the global eightfold rotational symmetry.
For given values of mean-fields, we numerically diagonalize
the mean-field Hamiltonian
$H_{MF}$ and update the mean-fields, and iterate this selfconsistent procedure
until the result converges within numerical accuracy.

\begin{figure}[htb]
 \centering
 \includegraphics[width=\linewidth]{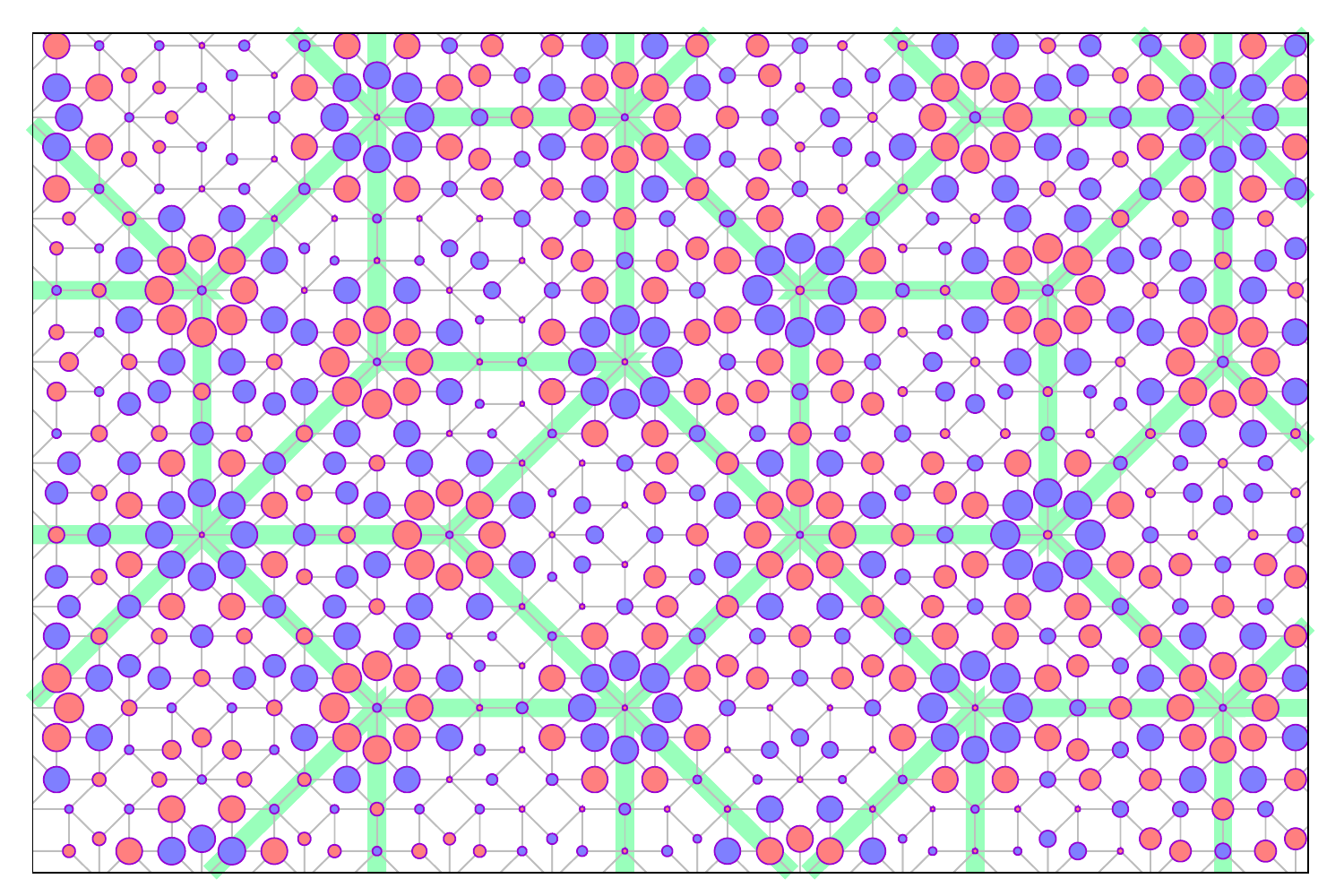}
 \caption{
   Spatial pattern for the staggered magnetization
   in the Hubbard model on the Ammann-Beenker tiling
   when $U/t=1.0\times 10^{-7}$ (essentially the same as $U=0$).
   The area of the circles represents
   the magnitude of the local magnetization.
   Bold lines represent the Ammann-Beenker tiling
   with the lattice constant $\tau^2$.
 }
 \label{dis0}
\end{figure}

We show in Fig.~\ref{dis0} the spatial pattern of the magnetization
$m_i(=\langle n_{i\uparrow}\rangle - \langle n_{i\downarrow}\rangle)/2$
when $U/t=1.0\times 10^{-7}$.
It is found that finite staggered magnetizations are induced even in the limit.
This is due to the existence of the confined states,
as discussed above.
We note that the F vertices are also magnetized
except for the F vertex at the center of the system.
This originates from the fact that
the amplitude of the confined states at the F$_n$ vertex is zero
in the $D_n$ domain, while should be finite in the larger domains,
discussed before.
Therefore, it is naively expected that, in the thermodynamic limit,
each lattice site have a finite magnetization even in the weak coupling limit.
This is in contrast to the systems with delta-function peak in DOS such as
the Lieb and Penrose lattices,
where there exist a finite density of nonmagnetic sites.
A remarkable point is that eight A and B vertices around the F vertex
have large magnetizations with $m\sim 1/16$ and
the other A and B vertices are less magnetized,
as shown in Fig.~\ref{dis0}.
Then, the Ammann-Beenker tiling with the larger lattice constant $\tau^2$ is formed
in the spatial distribution of the magnetizations
if the F vertex and adjacent A and B vertices with large magnetizations
are regarded as its ``unit cell''.
This may imply the superlattice structure (fractal behavior)
in the magnetic profile, which will be discussed later.

\begin{figure}[htb]
  \centering
  \includegraphics[width=\linewidth]{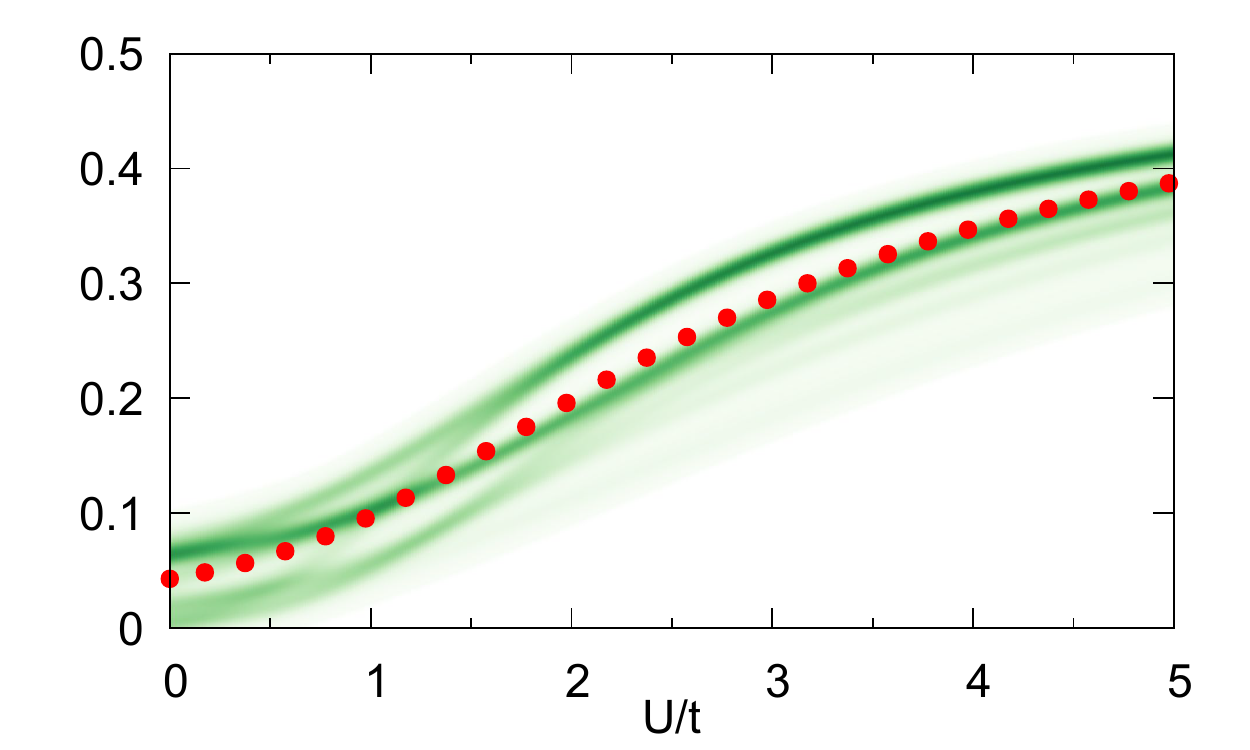}
  \caption{
   Distribution of local magnetizations as a function of the Coulomb interaction
   $U/t$ in the system with $N=180\,329$.
   Dotted lines represent the sublattice average $\bar{m}$.
 }
 \label{cross}
\end{figure}

Increasing the Coulomb interactions, the magnetizations monotonically increase
and finally the system should be described by the Heisenberg model
in the strong coupling limit.
To clarify the crossover in the ordered state
between weak and strong coupling regimes,
we show in Fig.~\ref{cross} the distribution of local magnetizations
in the system with $N=180\,329$.
When $U/t\rightarrow 0$,
a finite distribution appears in the magnetization,
where the average of the staggered magnetization $\bar{m}_0\sim 0.043$.
This originates from the existence of the macroscopically degenerate states
discussed above
and the staggered magnetization should be given as $1/4\tau^2$
in the thermodynamic limit.
The increase of the Coulomb interactions monotonically increases
the absolute value of local magnetization $\bar{m}_i\sim \bar{m}_{i0}+c_i U$,
where $\bar{m}_{i0}$ is the local magnetization at $U\rightarrow 0$ and
$c_i$ is the constant.
This $U$ dependence differs from that in the conventional bipartite
system, where the staggered magnetization usually increases as
$m\sim\exp(-a/U)$, with $a$ is constant.
On the other hand, this behavior is common to that
in the bipartite systems with the macroscopically degenerate states
at the Fermi level such as the Lieb~\cite{Noda} and
Penrose~\cite{Koga_Tsunetsugu_2017} lattices.
Increasing the interaction strength, the distribution of local magnetizations
gradually changes.
At last, when $U/t\gtrsim 2$, the magnetizations are classified by some peaks.
This classification is closely related to the coordination number
for each site,
which is different from the weak coupling case.
Therefore, the crossover occurs in the antiferromagnetically ordered state
around $U/t\sim 1.5$.
Namely, in the strong coupling regime, the larger magnetization appears in the A vertices
with smaller coordinations.
This should be consistent with the quantum Monte Carlo results
for the Heisenberg model~\cite{Wessel_Jagannathan_Haas_2003,Jagannathan_2005}
although the mean-field treatment cannot take into account
quantum fluctuations originating from intersite
correlations.

\begin{widetext}

  \begin{figure}[htb]
  \centering
  \includegraphics[width=\linewidth]{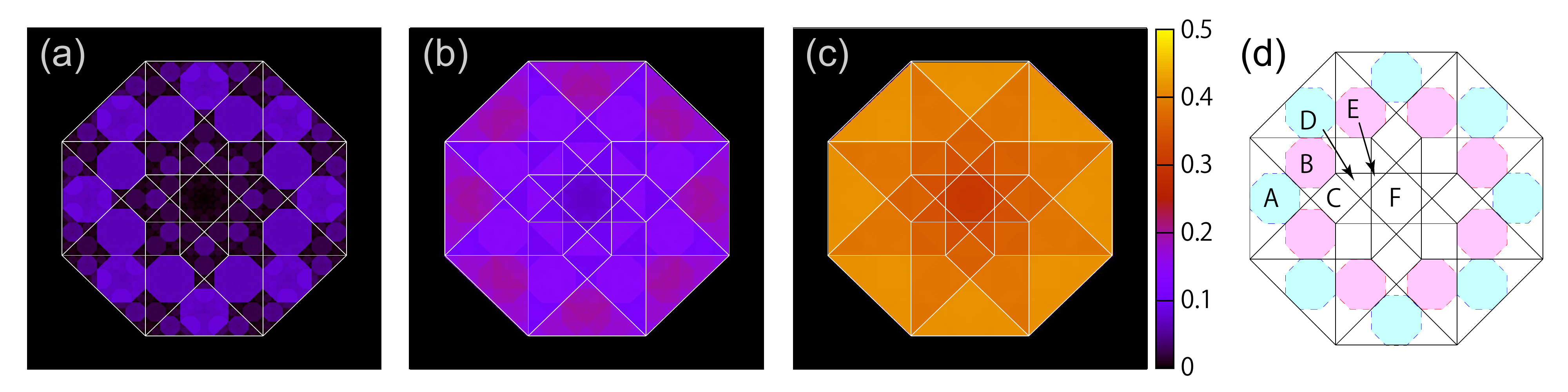}
  \caption{
    Magnetization profile in the perpendicular space $(\tilde{x}, \tilde{y})$
    for the Hubbard model
    when (a) $(U/t,N)=(1.0\times 10^{-7}, 1\,049\,137)$,
    (b) $(1.5,180\,329)$, and (c) $(5.0,180\,329)$.  
    (d) each area bounded by the solid lines
    is the region of one of the six types of vertices
    shown in Fig.~\ref{ABlattice}.
    Shaded areas bounded by the dashed lines
    represent the regions of certain A and B verties,
    which are the nearest-neigherbor and next-nearest-neighbor sites
    for the F vertices, respectively.
  }
 \label{perp}
\end{figure}
\end{widetext}

Finally, let us study the spatial profile
of the magnetizations characteristic of the Ammann-Beenker tiling.
To this end, we map it to the perpendicular space.
The positions in the perpendicular space have one-to-one correspondence
with the position in the physical space.
Each vertex site in the Ammann-Beenker tiling is described by
the four dimensional lattice points ${\vec n}=(n_0, n_1, n_2, n_3)$
labeled with integers $n_m$ (see Fig.~\ref{ABlattice}).
Their coordinates are the projections
onto the two-dimensional space:
\begin{eqnarray}
  {\bf r}&=&(x,y)=({\vec n}\cdot{\vec e}^x, {\vec n}\cdot{\vec e}^y),
\end{eqnarray}
where $e_m^x=\cos(m\pi/4)$ and $e_m^y=\sin(m\pi/4)$.
The projection onto the two-dimensional perpendicular space has
information specifying the local environment of each site,
\begin{eqnarray}
  {\bf {\tilde{r}}}=(\tilde{x},\tilde{y})=
    ({\vec n}\cdot\tilde{{\vec e}}^x, {\vec n}\cdot\tilde{\vec e}^y),
\end{eqnarray}
where $\tilde{e}_m^x=\cos(3m\pi/4)$ and $\tilde{e}_m^y=\sin(3m\pi/4)$.
Namely, six kinds of vertices have the corresponding regions
in the perpendicular space, as shown in Fig.~\ref{perp}(d).
Since vertices in both sublattices are uniformly distributed
in the corresponding regions of the perpendicular space,
the absolute value of magnetization are shown in Fig.~\ref{perp}.
In the weak coupling limit, we find the detailed structure in the perpendicular space,
meaning that the magnetization is not classified by the kinds of vertices.
Therefore, this magnetic profile is reflected by the spatial stucture of
the macroscopically degenerate confined states,
where large magnetizations appear in the A and B vertices around the F vertices,
as shown in Fig.~\ref{perp}(a).
Increasing the Coulomb interactions, interesting detailed structures smear
in the perpendicular space.
When $U/t=5$, the magnetizations are almost specified by
the vertices, where large magnetization appears in the A vertices
and small magnetization appears in the F vertices.
This tendency should be consistent with the results obtained
from the quantum Monte Carlo simulations~\cite{Wessel_Jagannathan_Haas_2003},
as mentioned above.

\begin{figure}[htb]
  \centering
  \includegraphics[width=\linewidth]{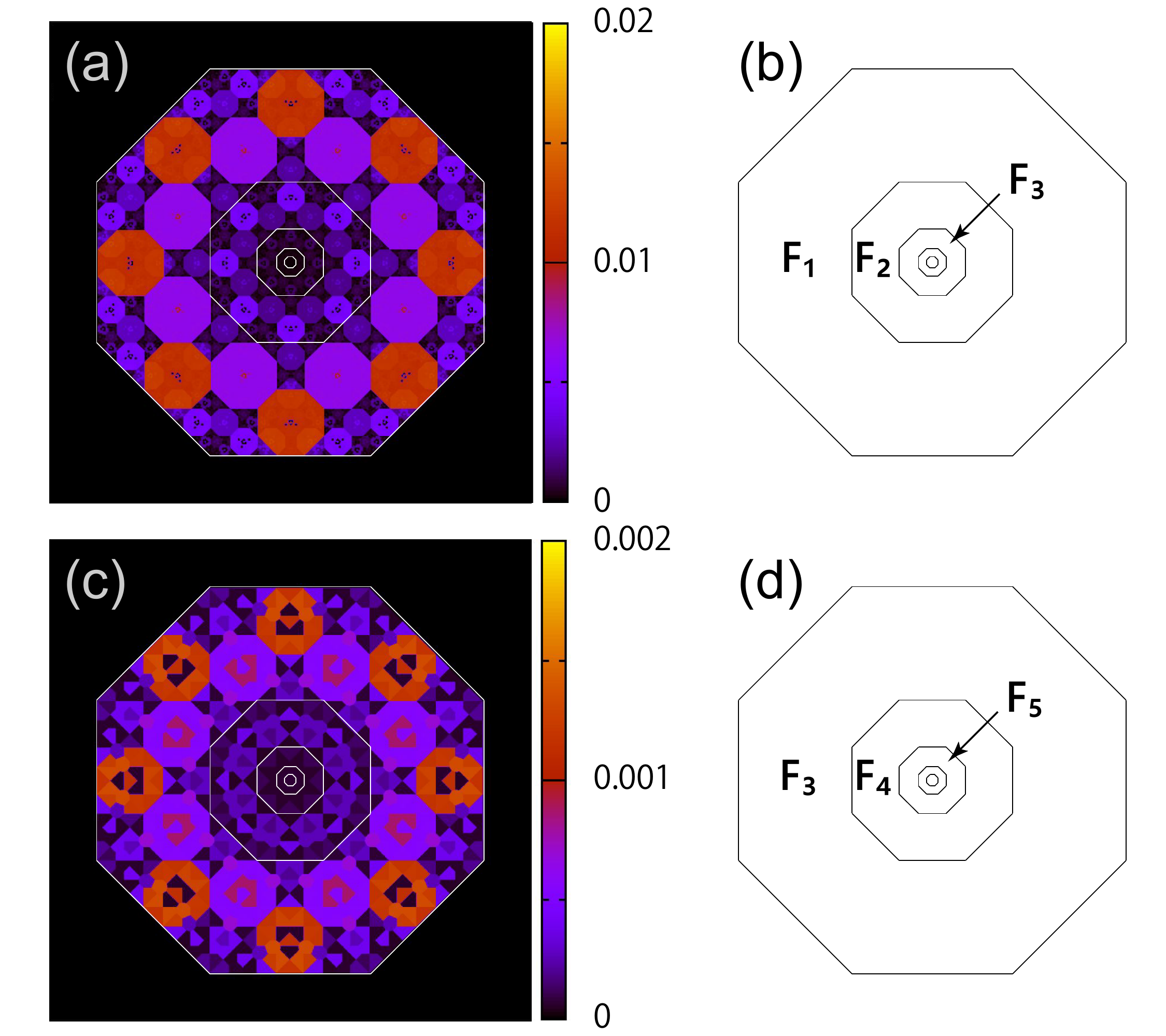}
  \caption{
    Magnetization profile in the perpendicular space
    for the system with $N=1\,049\,137$ when $U/t=1.0\times 10^{-7}$
    (essentially the same as 0).
    The results for the F [F$_i\; (i\ge 3)$] vertices are shown in (a) [(c)],
    and (b) [(d)] each part is the region of $F_i$ vertices.
  }
 \label{Fperp}
\end{figure}

Before summary, we wish to comment on fractal behavior
in the magnetic properties in the weak coupling case.
In the spatial distribution,
A and B vertices around the F vertex have large magnetizations and
these units form the Ammann-Beenker tiling with the lattice constant $\tau^2$,
as shown in Fig.~\ref{dis0}.
This superlattice structure in the magnetizations
allows us to consider the perpendicular space for the F vertex lattice.
Figure~\ref{Fperp}(a) shows the magnetization profile for the F vertices
in the weak coupling limit,
which is the same as that of the F vertex part in Fig.~\ref{perp}(a).
The average of the staggered magnetizations for the F vertices
$\bar{m}_0^{\rm F}\sim 0.005$
are much smaller than its bulk average $\bar{m}_0\sim 0.043$,
and therefore the magnetic profile for the F vertices
may be invisible in Fig.~\ref{perp}(a).
Figure~\ref{Fperp}(a) clearly shows that the magnetizations are not classified
by the kinds of the F vertices (F$_n$),
which are octagonally distributed in the perpendicular space,
as shown in Fig.~\ref{Fperp}(b).
Instead, we find the detailed structure in the distribution,
where ``A'' and ``B'' vertices around ``F'' vertex have large magnetization
in the Ammann-Beenker tiling with the lattice constant $\tau^2$.
This is similar to that in the original lattice shown in Fig.~\ref{perp}(a).
Therefore, we can say that a similar magnetic profile is found in this scale.
This may expect a further nested structure in the perpendicular space.
Considering the F$_i\; (i\ge 3)$ vertex lattice in the Ammann-Beenker
tiling with the lattice constant $\tau^4$,
we show the magnetic profile in their perpendicular space in Fig.\ref{Fperp}(c).
We find a similar detailed structure in the staggered magnetizations
although the number of the corresponding vertices are not large enough
and the absolute value of the magnetization is much smaller.
Then, we can say that fractal behavior appears in the magnetization profile,
in particular, in the weak coupling limit.

\section{Summary}
We have investigated magnetic properties in the half-filled Hubbard model
on the Ammann-Beenker tiling.
Considering the domain structure with locally eightfold rotational symmetry,
we have examined the strictly localized confined states.
We have then obtained their exact fractions in the thermodynamic limit. 
In contrast to the vertex model on the Penrose tiling,
the wave functions for confined states are densely distributed in the lattice
and thereby the introduction of the Coulomb interactions should induce
finite staggered magnetizations in each site.
Increasing the interaction strength, the spatial distribution of the magnetizations
continuously changes to those of the Heisenberg model.
Mapping the magnetization profiles to the perpendicular space,
we have clarified that
the superlattice structure appears in the magnetization profiles.

\begin{acknowledgments}
  We would like to thank Y. Takeuchi for fruitful discussions.
  Parts of the numerical calculations are performed
  in the supercomputing systems in ISSP, the University of Tokyo.
  This work was supported by Grant-in-Aid for Scientific Research from
  JSPS, KAKENHI Grant Nos.
  JP19H05821, JP18K04678, and JP17K05536.
\end{acknowledgments}

\section*{appendix}
Here, we prove that the number of the $\alpha$ vertex is
independent of the spin.
In the main text, we have proved that the numbers of squares and rhombuses are
independent of the spin, $S_\sigma=S/2$ and $R_\sigma=R/2$.
Now, we consider the inflation-deflation process for
the vertices~\cite{Socolar_1989,Baake_1990}.
Each vertex with the spin $\sigma$ is transformed
under the inflation process as,
\begin{equation}
  \begin{array}{lll}
    \begin{array}{rcl}
      A_\sigma&\rightarrow&0\\
      B_\sigma&\rightarrow&0\\
      C_{1\sigma}&\rightarrow&0\\
      C_{2\sigma}&\rightarrow&A_\sigma\\
      D_\sigma&\rightarrow&B_\sigma
    \end{array}
    &\;\;\;&
    \begin{array}{rcl}
      E_\sigma&\rightarrow&C_{1\sigma}\\
      F_\sigma&\rightarrow&\left\{
      \begin{array}{l}
        C_{2\sigma}\\
        D_\sigma\\
        E_\sigma\\
        F_\sigma
      \end{array}
      \right.,
    \end{array}
  \end{array}
\end{equation}
where 0 means that the vertices vanish under the inflation process.
Since there are two kinds of the C vertices in the tiling,
we have introduced C$_1$ and C$_2$ vertices.
Under the deflation process, a C$_{1\sigma}$ vertex is not changed from any vertex,
but is generated inside of each square with spin $\bar{\sigma}$, $S_{\bar{\sigma}}$,
as shown in Fig.~\ref{def}.
Therefore $p^{{\rm C}_{1\sigma}}=p^{S_{\bar{\sigma}}}/\tau^2=1/2\tau^3$,
where the fraction of the squares with spin $\sigma$ is 
  $p^{S_\sigma}=S_\sigma/\sum_{\sigma'}(S_{\sigma'}+R_{\sigma'})=1/2\tau$.
Another C$_{2\sigma}$ vertex is always generated from the A$_\sigma$ vertex,
$p^{{\rm C}_{2\sigma}}=p^{{\rm A}_\sigma}/\tau^2$.
Note that C$_1$ and C$_2$ vertices always appear
as the nearest-neighbor pair in the tiling, as shown in Fig.~\ref{F}.
Therefore, we can say that C$_2$ vertex is also independent of spin.
Since C$_{2\sigma}$ vertex is always changed to the A$_\sigma$ (E$_\sigma$) vertex
under the inflation (deflation) process, 
$p^{{\rm A}_\sigma}=p^{\rm A}/2$ ($p^{{\rm E}_\sigma}=p^{\rm E}/2$). 
Two B$_\sigma$ vertices are generated inside of each square $S_\sigma$ and
each rhombus $R_\sigma$, as shown in Fig.~\ref{def}.
This implies that the fraction of the B vertex is independent of the spin,
$p^{{\rm B}_\sigma}=p^{\rm B}/2$.
All B$_\sigma$ vertices are changed to the D$_\sigma$ vertices under the deflation process.
Therefore, the D vertex is also independent and
immediately we find that the F vertices are also independent.
Then, we can say that all vertices are independent of spins.
Namely, the F$_i$ vertices are also independent
since the F$_{i\sigma}$ vertices are generated from the F$_{i-1,\sigma}$ vertices
and the F$_{1\sigma}$ vertices are changed from the C$_{2\sigma}$, D$_\sigma$,
and E$_\sigma$ vertices under the deflation process.

\bibliography{./refs}

\end{document}